\documentclass[a4paper]{article}
\usepackage{INTERSPEECH2021}

\usepackage{subcaption}
\usepackage{xcolor}
\usepackage{multirow}
\usepackage[colorlinks,urlcolor=black]{hyperref}

\usepackage{enumitem}
\setlist{nosep} % or \setlist{noitemsep} to leave space around whole list

\newcommand{\dvec}{\textit{d}-vector}
\newcommand{\ivec}{\textit{i}-vector}
\newcommand{\xvec}{\textit{x}-vector}

\title{Chronological Self-Training for Real-Time Speaker Diarization}
\name{Dirk Padfield, Daniel J. Liebling}
\address{Google Research, USA}
\email{padfield@google.com, dliebling@google.com}
\begin{document}
%\ninept
%
\maketitle
\begin{abstract}
Diarization partitions an audio stream into segments based on the voices of the speakers.
Real-time diarization systems that include an enrollment step should limit enrollment training samples to reduce user interaction time.
Although training on a small number of samples yields poor performance, we show that the accuracy can be improved dramatically using a chronological self-training approach.
We studied the tradeoff between training time and classification performance and found that 1 second is sufficient to reach over 95\% accuracy.
We evaluated on 700 audio conversation files of about 10 minutes each from 6 different languages and demonstrated average diarization error rates as low as 10\%.
\end{abstract}

\noindent\textbf{Index Terms}: Diarization, real-time, {\dvec}, self-training, classification, clustering

\section{Introduction}
\label{sec:intro}

Speaker diarization is the process of partitioning an input audio into segments that indicate ``who spoke when.''
It has a large range of applications such as speaker turn change detection, audio annotation, and speaker verification.
Speaker diarization approaches can roughly be divided into two categories: real-time and offline.
Real-time approaches continually assign audio frames to speakers while speech is ongoing, which is useful for applications such as video call captioning and real-time conversation transcription.
Offline approaches are more restrictive in that they require the entire audio sequence to be available at once, and they are useful for scenarios such as transcribing voicemail \cite{Rosenberg2001CallerID} and stored videos.

We employ real-time diarization to serve a practical purpose: to enable streaming diarization in a real-time conversation app developed by our team for research purposes. 
Dyadic conversations in the app begin on a screen with two buttons, one for each talker. Conversations proceed in 2 stages: in stage 1, each user speaks while they push their own button on the screen, collecting speech samples for model training.  The embedding vectors collected during this period are the features, and the labels come from whichever button the respective user pressed.  
In stage 2, the app indicates to the speakers that it is ready to continue their conversation without the need to push buttons to take turns. 
Since button-pushing is tedious, prone to timing errors, and not a natural conversational interaction, we aim to limit the training phase as much as possible.
Also, unlike many diarization approaches in the literature that have access to the audio and labels of the full conversation upfront, our app requires that 
all model inference be done on the device in a streaming fashion. 
To meet these needs, we developed an approach that performs highly accurate streaming diarization with as little training and user interactions as possible by utilizing a self-training approach that adapts over time.
We call this approach chronological self-training.

Most diarization systems are composed of multiple components such as a voice activity detector (VAD) that partitions the audio into speech regions, an embedding model that embeds sliding windows of speech audio into a high-dimensional space, and a clustering stage that clusters the embeddings by speaker. Fully end-to-end neural diarizers such as \cite{Fujita2019} use a single system without separate embeddings. For embedding-based systems, some approaches also include an additional post-processing step where the clusters are further refined and combined. For example, Sell and Garcia-Romero \cite{sell2014} used probabilistic linear discriminant analysis (PLDA) to refine {\ivec}s \cite{Dehak:11, Shum:13}. Popular choices of embeddings include {\ivec}s, {\dvec}s \cite{Variani:14}, or {\xvec}s \cite{xvector}. 
Both {\xvec}s and {\dvec}s are DNN embeddings and have similar performance, and we prefer to use {\dvec}s.
Wan et al. \cite{Wan:18} showed that {\dvec}s are more effective than {\ivec}s because they utilize neural networks which were trained on large datasets of speakers with varying accents and acoustic conditions. The {\dvec}s were used in a full diarization system \cite{Wang:18} which included a variety of clustering approaches.  Their naive clusterer and the Links \cite{Mansfield:18} clusterer both process the stream in real-time.
Higher-performing offline algorithms such as k-means or spectral clustering can be used if the clustering is performed after all the points have been collected. Singh and Ganapathy \cite{Singh2020} combined representation learning with agglomerative hierarchical clustering (AHC) for significant improvement over AHC alone.
A fully-neural offline diarization system can also use unsupervised clustering \cite{Zhang:19}.

Our main contributions are:
1) demonstrate a real-time diarization system with limited enrollment that achieves high accuracy and low diarization error rate,
2) present measurements of the tradeoff between training time and diarization accuracy in a real-time setting, and
3) provide large-scale quantification of diarization performance on 700 audio files across 6 different languages.

\section{Methods}
\label{sec:methods}
        
\begin{figure*}[hbtp]
    %\centering
    \begin{subfigure}[t]{0.48\linewidth}
        \centering
        \includegraphics[width=\textwidth]{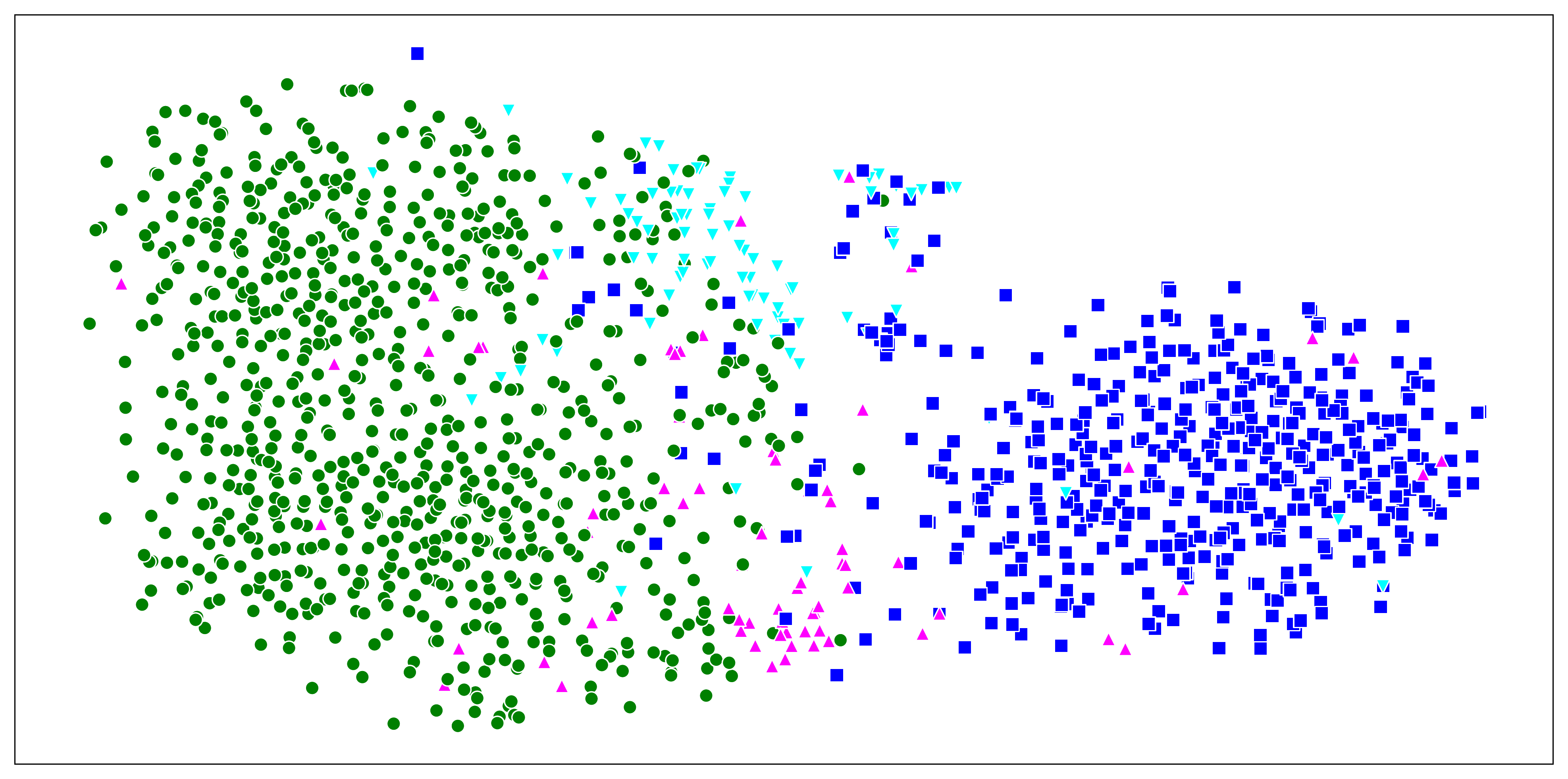}
        %en 5254
        \caption{}
        \label{fig:en_5254_tSNE_all_points}
    \end{subfigure}
    \hfill
    \begin{subfigure}[t]{0.48\linewidth}
        \centering
        \includegraphics[width=\textwidth]{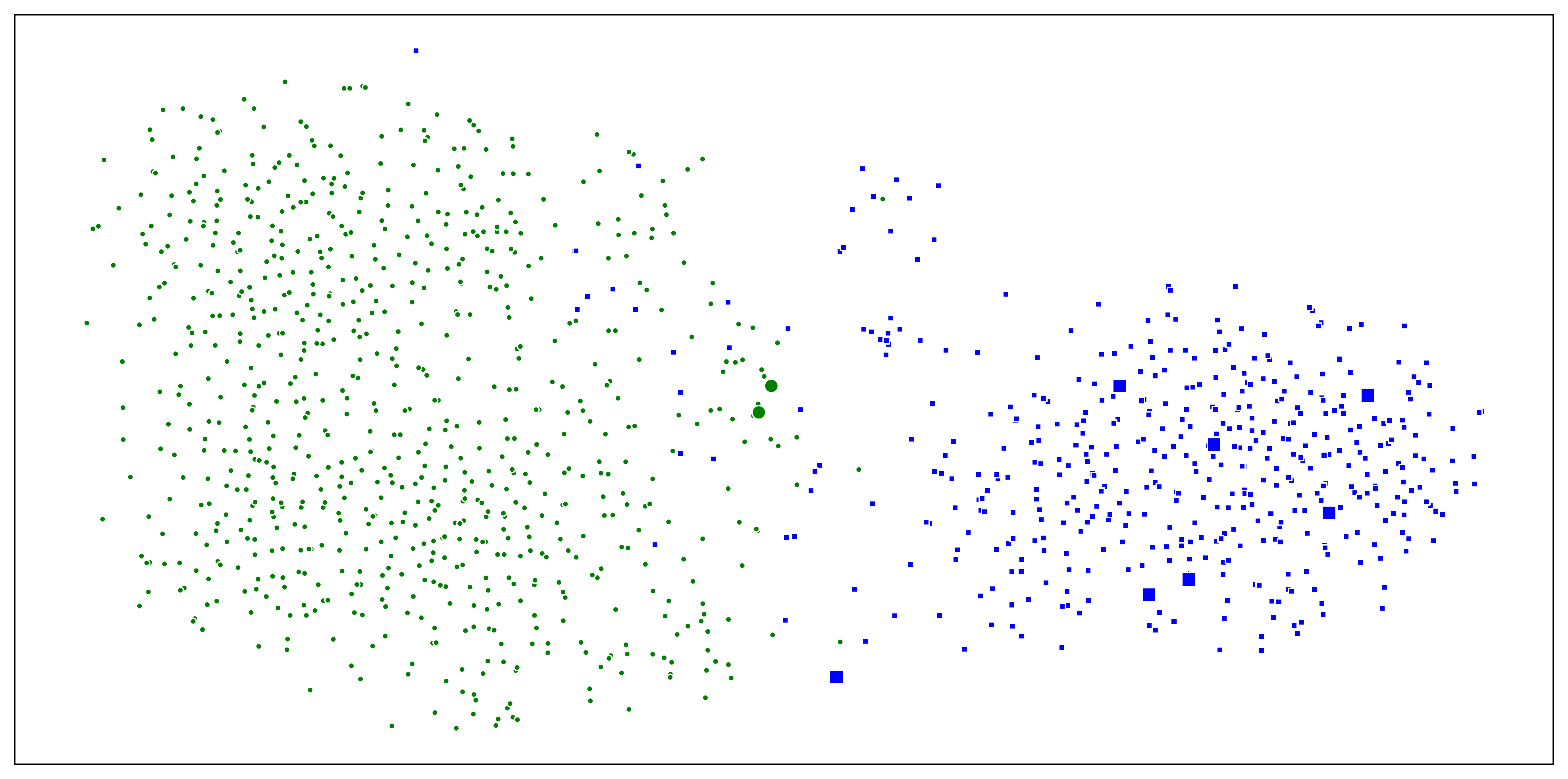}
        \caption{}
        \label{fig:en_5254_tSNE_training_points}
    \end{subfigure}
    \hfill
    \begin{subfigure}[t]{0.48\linewidth}
        \centering
        \includegraphics[width=\textwidth]{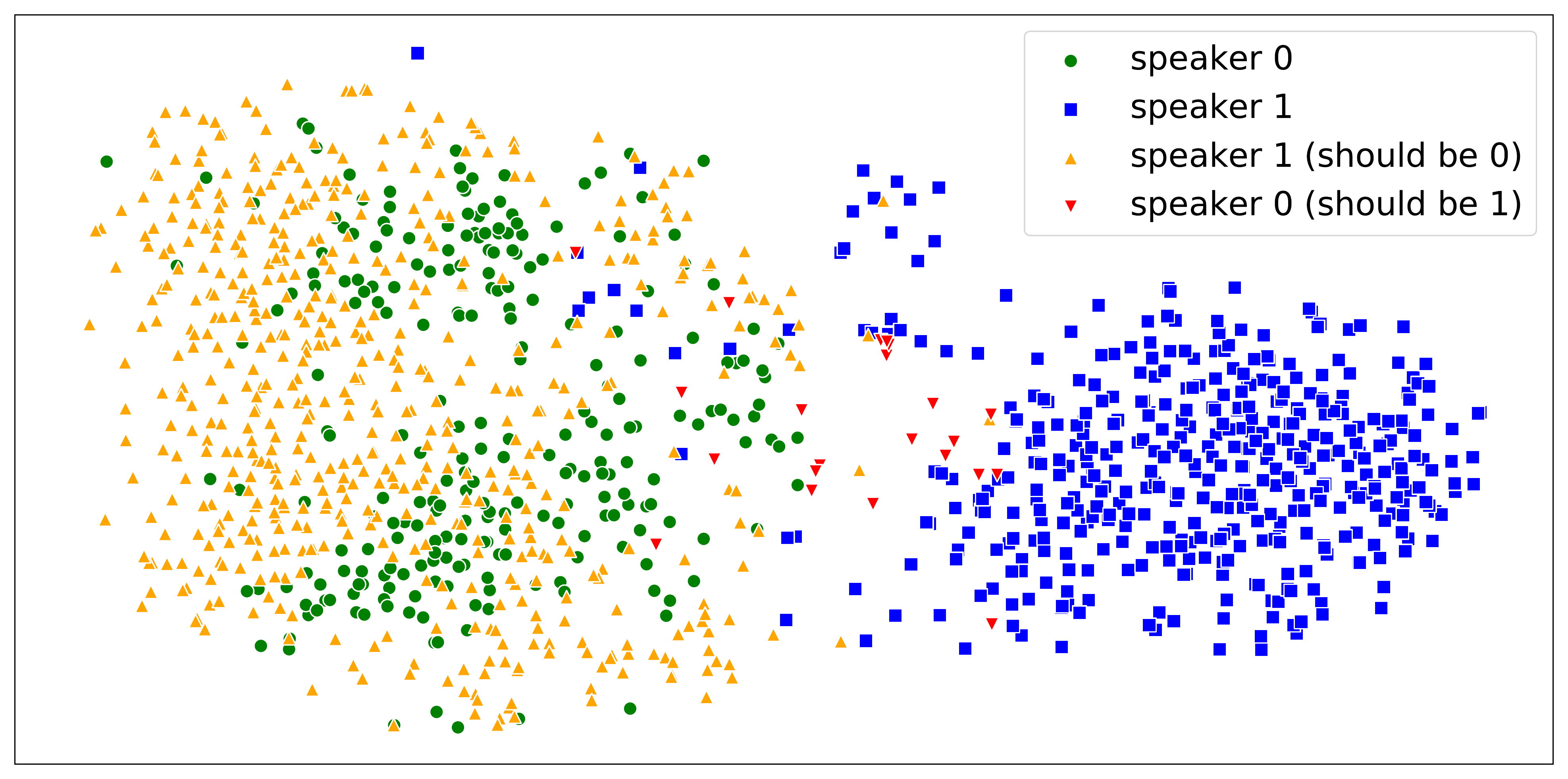}
        \caption{}
        \label{fig:en_5254_tSNE_NearestCentroid}
    \end{subfigure}
    \hfill
    \begin{subfigure}[t]{0.48\linewidth}
        \centering
        \includegraphics[width=\textwidth]{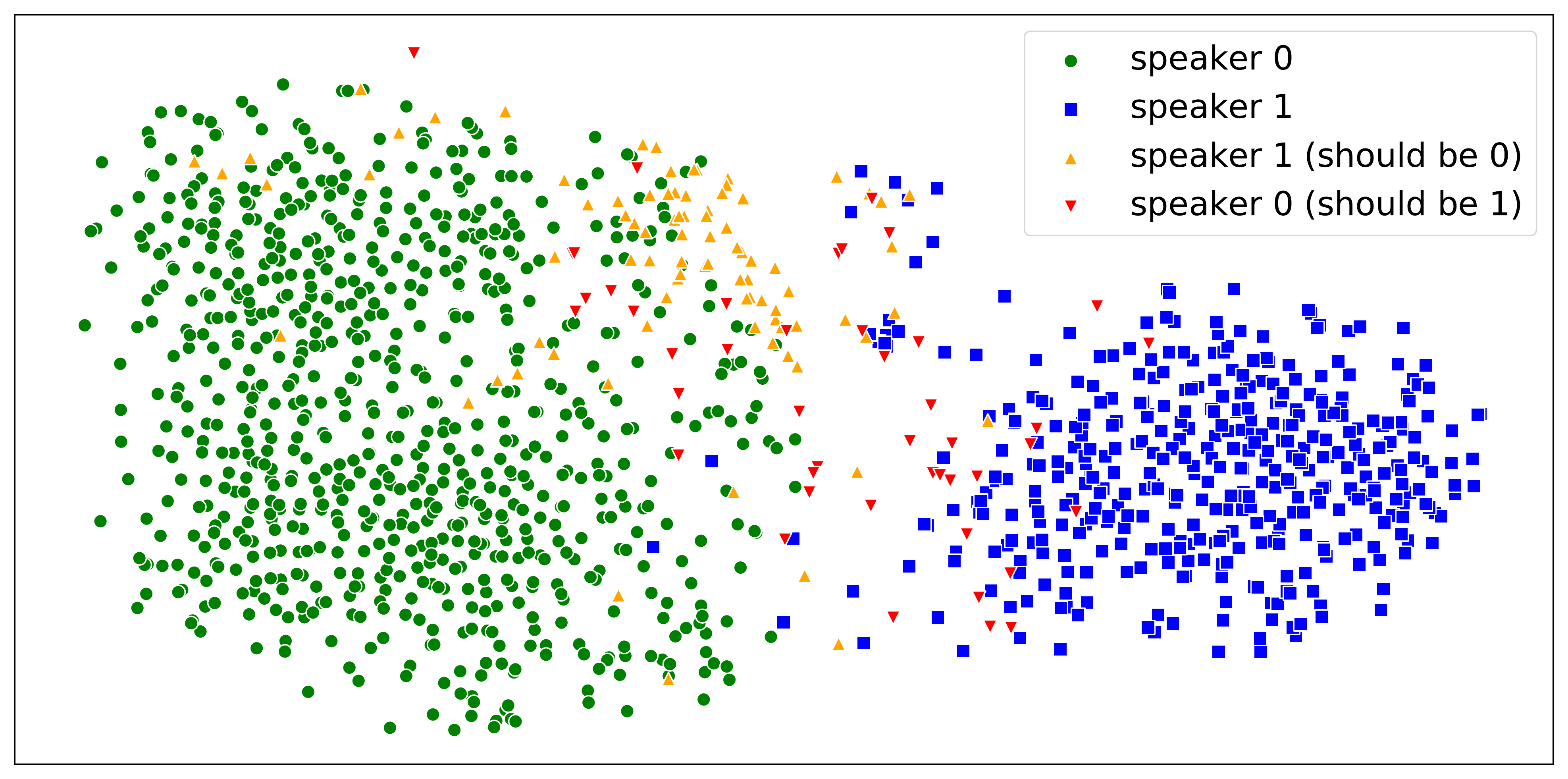}
        \caption{}
        \label{fig:en_5254_tSNE_AdaptiveNearestCentroid}
    \end{subfigure}
        \caption{Visualization of training points and predictions using 2D t-SNE projection of the 256-dimensional {\dvec}s of a conversation.
        (a) Blue is one speaker, green another, magenta is overlapping speech, and cyan is when no one was speaking. 
        (b) The large points are the first 10 samples of the conversation.  These are used for training while the rest are used for testing.
        (c) Classification using Nearest Centroid with an accuracy of 49\%.  The yellow and red points are mis-classifications.
        (d) Classification using adaptive Nearest Centroid with an accuracy of 85\%.  Many of the mis-classified points are corrected.
        }
        \label{fig:adaptation}
\end{figure*}

Our approach builds upon prior work on {\dvec} embeddings that embed a speaker's voice in a 256-dimensional hyper-sphere in such a way that the points of the same speaker are close and those of different speakers are separated \cite{Wan:18,Wang:18}.
Each audio file of conversations is passed through the embedding pipeline, which first partitions the audio into speech regions using a VAD, extracts embeddings from sliding windows across the speech regions, averages the embeddings, and computes {\dvec}s.
{\dvec}s are computed approximately every 200ms, which we found experimentally to be sufficiently frequent to capture changes in speaker turns while not being computationally intensive.  
More details such as architecture and training data are in \cite{Wang:18}.

The processing step yields sequences of {\dvec}s for the conversations, and the goal is to group these together into speaker turns.
Each {\dvec} is associated with the ground truth speaker IDs by comparing with the timestamps from human-annotated files.  
Many diarization approaches use a separate enrollment file recorded with different channel conditions to compare against during evaluation.  
However, our app is designed to work for any new speaker,
so we need to run an enrollment step at the start of each speaker session and then predict on the remainder of the session.

To model our application use-case, our approach divides the {\dvec}s into training and testing sets chronologically by time.
For each audio file, we train multiple classifiers on the {\dvec}s and their labels from the first $t$ seconds and test them on the remaining {\dvec}s.
By training and testing on each file independently, each model learns its parameters during the enrollment stage.
Although this ensures that the speaker and channel characteristics of the {\dvec} embeddings in the enrollment remain the same during testing, this is not an artificial setup because it perfectly emulates our scenario of the app being trained as users initially take turns.
It would be convenient to use an enrolment protocol that guarantees an equal representation of the speakers, but that would also not meet our use case.
To  accurately represent our use-case, we designed our approach to use the first $t$ seconds of a user's speech for training.  

Since our application is constrained to operate in real-time, our experimentation is confined to algorithms that can be trained on partial data and updated over time.
We used the following classifiers: 
1) K-Nearest Neighbors (K-NN), which labels test points using the mode of the labels of the $K$ closest samples in the training set,
2) Gaussian Na\"ive Bayes (GNB), which fits a Gaussian during training and computes the maximum likelihood estimate during testing, 
3) and Nearest Centroid (NC), which computes the centroid for each class during training and measures the distance to those centroids during testing.
We used the scikit-learn package \cite{scikit-learn} with the following settings: 
K-NN (metric=cosine, neighbors=3),
GNB (var smoothing=0.1),
and NC (metric=cosine). 
These algorithms rely on key assumptions of sample distributions that may not fully hold for {\dvec}s, but they perform well in practice as shown in Section \ref{sec:results}.

\subsection{Classification of embeddings}

Each embedding vector represents a time window of about 200ms, and we can visualize the separability of these embeddings by running t-SNE \cite{Hinton:03} to project the points onto a 2D plane while roughly maintaining their relative distances. 
In Figure \ref{fig:en_5254_tSNE_all_points}, after the dimensionality reduction with t-SNE, each point in a single, sample conversation is colored by its human label: blue and green are different speakers.
This is useful for visualizing the speaker points and the results of classification, but it cannot be used for classification itself because the projection cannot be re-used on new points once it has been computed.  
Thus the classification algorithms are run on the original 256-dimensional vectors.

In a real conversation, there is no control over which speaker will speak first and for how long, so the algorithm has to handle whatever audio it receives first. 
Thus, the classifiers only have access to the first set of samples of the conversation for training.
In Figure \ref{fig:en_5254_tSNE_training_points}, 
the chronologically first 10 points (shown as large dots) from a sample audio conversation are the only points the classifiers can use for training.
The remaining points (shown as smaller dots) are used for testing.  
In this example, only 2 of these samples come from the ``green'' speaker (2 large green dots), and these points lie on the far right border of the cloud of points and do not accurately represent the entire distribution.
This can happen if the enrollment audio clips do not accurately represent the full range of the speaker's voice and can lead to mis-classification of later points.

Such limited and poorly distributed training points have a dramatic impact on the classification performance as shown in Figure \ref{fig:en_5254_tSNE_NearestCentroid}, where a large swath of green points are mislabeled (shown in yellow).
This is reflected in the accuracy, which is only 49\% in this case.
One solution is to collect more training data, but this would result in more user interaction time.

\subsection{Chronological self-training}

To achieve higher performance without changing the training data, we allow the algorithm to adapt over time after its initial training stage instead of training simply on the training window.
We accomplish this with a self-training procedure \cite{Scudder:65, Yarowsky:95}, where the model is continually updated with its predictions on new unlabeled samples as they arrive:
\begin{enumerate}
    \item Train the model on the initial training points.
    \item \label{prediction_step} Predict the labels on the next set of $B$ test points, where $B$ represents a batch size (that we usually set to 10).
    \item Retrain the model including the predicted labels, optionally ignoring points with low prediction score.
    \item Return to step 2.
\end{enumerate}

If the batch size is too high, the model will adapt slowly, and if too low, the model will have to re-train too often. 
Using hyper-parameter tuning \cite{vizier}, we found that 10 is a good tradeoff.
We implemented this adaptive approach for both the NC and GNB algorithms, both of which have the flexibility to operate on a partial set of points.
Since each prediction step yields a probability for each test point, we could optionally limit the model update to points whose probabilities exceed a desired threshold,
but we found through experimentation that this led to only modest gains and that simply including all subsequent predicted labels to update the model yielded high-quality results.
The model could drift if too few training samples are included such that noisy predictions dominate; we present some analysis of this tradeoff in Figure \ref{fig:test_scores}.

Running the adaptive NC algorithm increases the accuracy of Figure \ref{fig:en_5254_tSNE_NearestCentroid} from 49\% to 85\% even though the training samples are the same as shown in Figure \ref{fig:en_5254_tSNE_training_points}.  
This can be seen in Figure \ref{fig:en_5254_tSNE_AdaptiveNearestCentroid}, where many more points are classified correctly.
Most of the remaining mistakes (yellow and red) fall on the border between the clusters, which is the most challenging region.
It is promising that the algorithm can reach high accuracy with such limited training data, and this impact is studied in Section \ref{sec:results}.

\section{Experimental Results}
\label{sec:results}

We used 700 CALLHOME (\href{https://catalog.ldc.upenn.edu}{https://catalog.ldc.upenn.edu}) audio files, each with 10 minutes of manually annotated audio, from 6 different languages:
English (\href{https://catalog.ldc.upenn.edu/LDC97S42}{LDC97S42}), 
Spanish (\href{https://catalog.ldc.upenn.edu/LDC96S35}{LDC96S35}), 
German (\href{https://catalog.ldc.upenn.edu/LDC97S43}{LDC97S43}), 
Arabic (\href{https://catalog.ldc.upenn.edu/LDC97S45}{LDC97S45}), 
Mandarin (\href{https://catalog.ldc.upenn.edu/LDC96S34}{LDC96S34}),
and Japanese (\href{https://catalog.ldc.upenn.edu/LDC96S37}{LDC96S37}).
This dataset consists of audio recordings of real phone conversations among 2 or more participants, and they closely match our scenario of real-time conversations.
Each conversation includes a rich set of manually-generated diarization annotations, which enabled us to evaluate our approach on this large dataset that closely resembles our streaming diarization use case.

For each audio file, we merged multiple channels into a single channel as is typical in the literature \cite{Sell:15, Shum:13}.
In each audio file, we use the un-partitioned evaluation map (UEM) files to determine the region of human annotation and do not process the parts that are before the first annotation or after the last annotation.
DER is the main comparison metric and is calculated using the pyannote.metrics library \cite{Bredin2017}.
When computing the metrics, we exclude overlapped speech (multiple speakers speaking at the same time) and tolerate errors less than 250ms (collar = 0.25) in locating segment boundaries as is the standard convention in the literature \cite{Sell:15, Shum:13, Castaldo:08, Senoussaoui:14, Garcia-Romero:17}.

Since our experimental setup relies on the manual annotations provided in the CALLHOME data, there is no need for manual annotation of individual d-vectors.
For the training data, our system selects the {\dvec}s in the first $t$ seconds of each audio file along with the manual annotations.
This enables us to avoid the problem of having to manually annotate the very short audio segments corresponding to the {\dvec}s, which are too short for humans to listen to and annotate in isolation. 

\subsection{Influence of training time on accuracy}

Each speaker enrolls for a given number of seconds, and we aim to reduce this while retaining high accuracy.
We measured the tradeoff between training time and performance by computing the performance while sweeping the training time per speaker in steps of 1 second from 1 to 10 (and one at 0.5 seconds).
Users will realistically not train the system for more than 10 seconds, and all of the algorithms performed well with just a few seconds of training data.
At each setting, the script accumulates the time spoken by each person in an audio file until the required time is reached, which marks the end of the training samples. 
For consistency, the testing samples were held fixed across all settings by starting at the first sample after the last training sample across tests. 

\begin{figure}[htb]
    \centering
    \includegraphics[width=\linewidth]{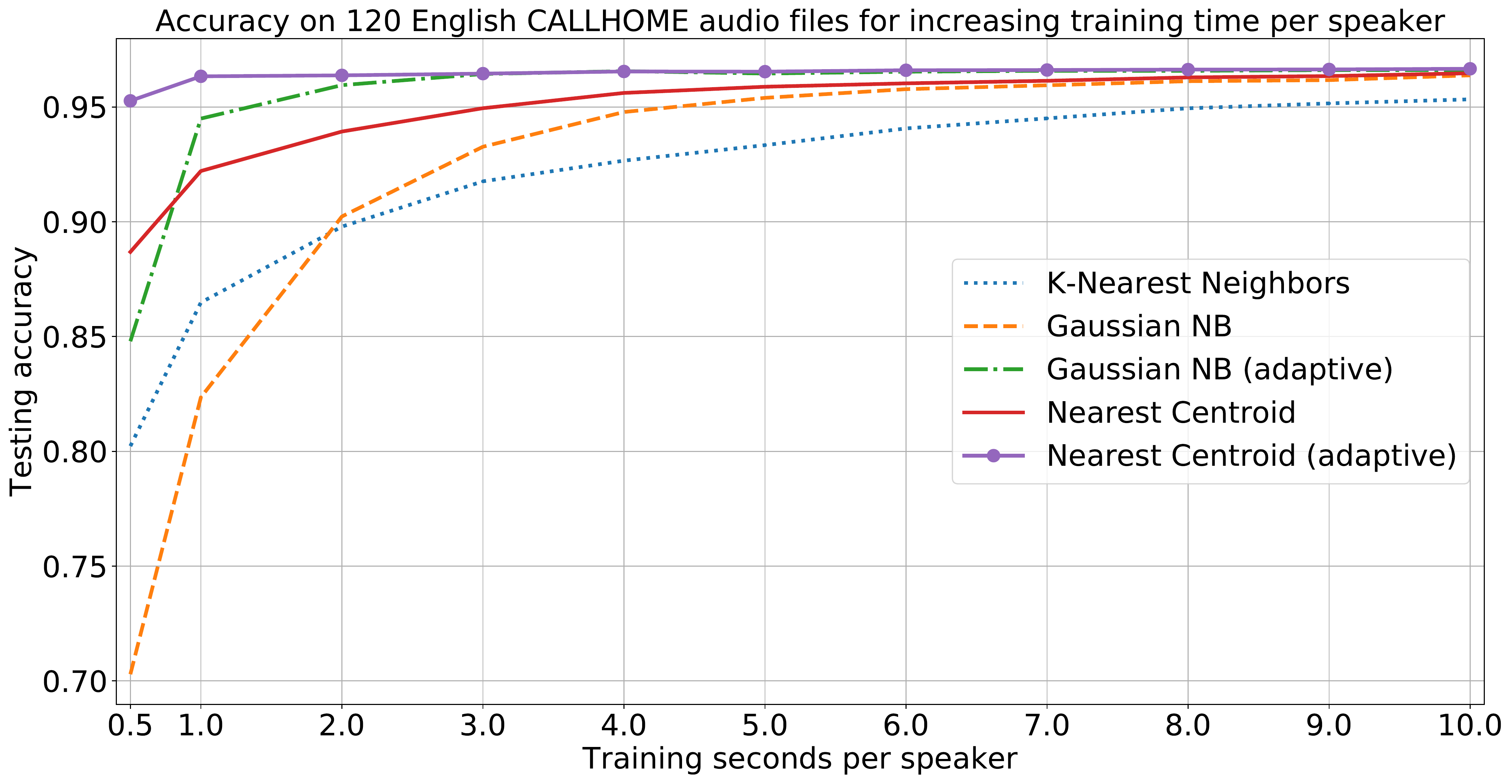}
    \caption{Mean testing accuracy for the {\dvec}s of each classifier as a function of training time.}
    \label{fig:test_scores}
\end{figure}

Figure \ref{fig:test_scores} shows plots of the mean accuracy of each algorithm over the 120 audio files of the English dataset as a function of training time.
This plot shows that the adaptive NC and adaptive Naive Bayes methods saturate very early, which  demonstrates the strength of the adaptive approach: it reaches peak performance with very little training data.  
The reason why adding more than 3 seconds does not provide additional gains is that there is human error in the annotations so that the classifiers cannot reach 100\%; we estimated about a 3\% error in the human annotations.

The accuracy metric is used because the speakers are mostly balanced in each conversation.
The adaptive NC algorithm performs the best, reaching over 95\% accuracy with just 0.5 seconds (2 to 3 {\dvec} samples) of training time.
The adaptive GNB, NC, GNB, and K-NN classifiers require about 1, 3, 4, and 8 seconds, respectively, to reach this level of accuracy.  
The standard deviation of accuracy across audio files was also computed but is not shown so as not to clutter the plots.
Adaptive NC has the smallest standard deviation (for example, 3.2 at 1 second), and the variance decreases with training time for all of the algorithms.

\subsection{Quantitative comparisons}
\label{sec:quantitative}

We next computed the diarization error rate (DER) to compare the performance of the classification algorithms.
Table \ref{tab:results} provides detailed scores for multiple algorithms on the 700 audio files, each about 10 minutes long from 6 languages: English (EN), Spanish (SP), German (DE), Arabic (AR), Mandarin (CN), and Japanese (JP).
The scores for each language represent the average of all the audio files for that language.

\begin{table}[htbp]
\footnotesize
\begin{centering}
\caption{
Average error rates for CALLHOME audio ($n=700$), grouped by language. 
FA = False Alarm, Miss = missed detection.
Bold text indicates the best DER per language.
The processing pipeline and settings were kept constant across all experiments.
}
\begin{tabular}{|l|l|c|c|c|c|}
\hline
 & Method & Confusion & FA & Miss & DER \\
\hline\hline
\multirow{8}{*}{EN} & Naive & 22.85 & \multirow{8}{*}{2.09} & \multirow{8}{*}{4.55} & 29.49\\
& Links & 16.38 &  &  & 23.03 \\
& Spectral & 5.43 &  &  & 12.07 \\
& K-NN & 11.74 &  &  & 18.38 \\
& GNB & 15.14 &  &  & 21.78 \\
& GNB(adapt) & 4.78 &  &  & 11.42 \\
& NC & 6.81 &  &  & 13.45 \\
& NC(adapt) & {\bf 3.31} &  &  & {\bf 9.95} \\
\hline\hline
\multirow{8}{*}{SP} & Naive & 26.75& \multirow{8}{*}{0.84} & \multirow{8}{*}{8.20} & 35.79\\
& Links & 20.31 &  &  & 29.35 \\
& Spectral & {\bf 7.28} &  &  & {\bf 16.32} \\
& K-NN & 15.39 &  &  & 24.43 \\
& GNB & 17.39 &  &  & 26.43 \\
& GNB(adapt) & 9.02 &  &  & 18.06 \\
& NC & 10.90 &  &  & 19.93 \\
& NC(adapt) & 8.48 &  &  & 17.51 \\
\hline\hline
\multirow{8}{*}{DE} & Naive &22.09 & \multirow{8}{*}{0.76} & \multirow{8}{*}{6.15} & 29.00 \\
& Links & 15.95 &  &  & 22.85 \\
& Spectral & 3.96 &  &  & 10.87 \\
& K-NN & 9.18 &  &  & 16.09 \\
& GNB & 12.91 &  &  & 19.82 \\
& GNB(adapt) & 4.73 &  &  & 11.63 \\
& NC & 5.96 &  &  & 12.86 \\
& NC(adapt) & {\bf 3.83} &  &  & {\bf 10.73} \\
\hline\hline
\multirow{8}{*}{AR} & Naive & 28.69& \multirow{8}{*}{0.65} & \multirow{8}{*}{6.39} & 35.73 \\
& Links & 21.37 &  &  & 28.41 \\
& Spectral & 8.71 &  &  & 15.75 \\
& K-NN & 16.79 &  &  & 23.82 \\
& GNB & 19.14 &  &  & 26.18 \\
& GNB(adapt) & 9.90 &  &  & 16.93 \\
& NC & 12.25 &  &  & 19.28 \\
& NC(adapt) & {\bf 8.61} &  &  & {\bf 15.65} \\
\hline\hline
\multirow{8}{*}{CN} & Naive & 27.32& \multirow{8}{*}{1.96} & \multirow{8}{*}{4.38} & 33.66 \\
& Links & 19.77 &  &  & 26.10 \\
& Spectral & {\bf 8.11} &  &  & {\bf 14.44} \\
& K-NN & 17.37 &  &  & 23.70 \\
& GNB & 19.80 &  &  & 26.13 \\
& GNB(adapt) & 9.90 &  &  & 16.23 \\
& NC & 11.91 &  &  & 18.24 \\
& NC(adapt) & 9.46 &  &  & 15.79 \\
\hline\hline
\multirow{8}{*}{JP} & Naive & 25.19 & \multirow{8}{*}{1.07} & \multirow{8}{*}{9.96} & 36.22 \\
& Links & 20.03 &  &  & 31.06 \\
& Spectral & {\bf 6.06} &  &  & {\bf 17.09} \\
& K-NN & 13.14 &  &  & 24.17 \\
& GNB & 15.44 &  &  & 26.47 \\
& GNB(adapt) & 6.94 &  &  & 17.97 \\
& NC & 9.00 &  &  & 20.03 \\
& NC(adapt) & 6.40 &  &  & 17.43 \\
\hline
\end{tabular}
\label{tab:results}
\end{centering}
\end{table}

The DER is composed of the confusion (mis-classified speakers), false alarm (FA: speaker detected when none present), and missed detection (Miss: true speech not detected).
Within each language, the FA and Miss are the same for all algorithms since they depend only on the voice activity detector (VAD), whose settings are held constant across all experiments.
Only the confusion is impacted by the algorithms.

Both classification and clustering algorithms are presented in the table.
For each language, the first 3 algorithms are unsupervised clustering algorithms, and the last 5 are classifiers.
The clustering algorithms are the Naive, Links, and Spectral algorithms from \cite{Wang:18}.
The Naive and Links algorithms run in real-time, updating their clusters as new points are added, but their DER is often too high to be usable in practice as shown in Table \ref{tab:results}.
The Spectral algorithm works offline, requiring all points to be present, and thus it is not viable for real-time processing, but it does serve as an upper-bound on clustering performance.
For all experiments, we used the same experimental settings, diarization metrics, and processing pipeline as those used in \cite{Wang:18} for their CALLHOME results with the exception that we used a slightly updated embedding model and we evaluated on all audio files for each language instead of a subset.
Thus, our results for the Naive, Links, and Spectral algorithms are comparable to the results in \cite{Wang:18}, which themselves were shown to outperformed those in \cite{Shum:13, Sell:15, Castaldo:08, Senoussaoui:14, Garcia-Romero:17}.

In all experiments, all the classifiers were trained with only 1 second of audio per speaker.
The adaptive NC performs the best across the {\it classification} algorithms and performs on-par with the Spectral clusterer across all languages.
This is true even though the classifier runs in real-time whereas the Spectral clusterer has access to all the data at once (although without any training labels).
This demonstrates the effectiveness of enabling the model to chronologically self-train on test samples without providing it with the labels of those test samples.

\section{Conclusions}
\label{sec:conclusions}

Our goal is to build accurate real-time diarization algorithms requiring as little human input for enrollment as possible.
Training on a small amount of initial audio from speakers is usually not enough to fully characterize their voice, but chronological self-training can adapt on test data and dramatically improve the performance.
The adaptive NC classifier with only 1 second of training data per speaker yields the lowest DER rates of all classifiers on the CALLHOME datasets.

For future work, 
we plan to investigate the possibility of treating the clustering as a tracking problem as the embedding points move on the hyper-sphere.
We also plan to investigate the distribution of the error over time to look for patterns such as whether the majority of errors come from the samples that are distant in time from the training examples.

\section{Acknowledgements}
We want to thank Dick Lyon, Audrey Addison, and Ramin Mehran for their helpful manuscript suggestions.

\bibliographystyle{IEEEtran}
\bibliography{references}

\begin{thebibliography}{10}
\providecommand{\url}[1]{#1}
\csname url@samestyle\endcsname
\providecommand{\newblock}{\relax}
\providecommand{\bibinfo}[2]{#2}
\providecommand{\BIBentrySTDinterwordspacing}{\spaceskip=0pt\relax}
\providecommand{\BIBentryALTinterwordstretchfactor}{4}
\providecommand{\BIBentryALTinterwordspacing}{\spaceskip=\fontdimen2\font plus
\BIBentryALTinterwordstretchfactor\fontdimen3\font minus
  \fontdimen4\font\relax}
\providecommand{\BIBforeignlanguage}[2]{{%
\expandafter\ifx\csname l@#1\endcsname\relax
\typeout{** WARNING: IEEEtran.bst: No hyphenation pattern has been}%
\typeout{** loaded for the language `#1'. Using the pattern for}%
\typeout{** the default language instead.}%
\else
\language=\csname l@#1\endcsname
\fi
#2}}
\providecommand{\BIBdecl}{\relax}
\BIBdecl

\bibitem{Rosenberg2001CallerID}
\BIBentryALTinterwordspacing
A.~E. Rosenberg, J.~Hirschberg, M.~Bacchiani, S.~Parthasarathy, P.~L. Isenhour,
  and L.~Stead, ``Caller identification for the {SCANMail} voicemail browser,''
  in \emph{Proceedings of Eurospeech 2001}, 2001. [Online]. Available:
  \url{https://www.isca-speech.org/archive/eurospeech_2001/e01_2373.html}
\BIBentrySTDinterwordspacing

\bibitem{Fujita2019}
\BIBentryALTinterwordspacing
Y.~Fujita, N.~Kanda, S.~Horiguchi, K.~Nagamatsu, and S.~Watanabe, ``End-to-end
  neural speaker diarization with permutation-free objectives,'' in
  \emph{Proceedings of Interspeech 2019}, 2019, pp. 4300--4304. [Online].
  Available: \url{http://dx.doi.org/10.21437/Interspeech.2019-2899}
\BIBentrySTDinterwordspacing

\bibitem{sell2014}
\BIBentryALTinterwordspacing
G.~Sell and D.~{Garcia-Romero}, ``Speaker diarization with {PLDA} i-vector
  scoring and unsupervised calibration,'' in \emph{2014 IEEE Spoken Language
  Technology Workshop (SLT)}, 2014, pp. 413--417. [Online]. Available:
  \url{https://doi.org/10.1109/SLT.2014.7078610}
\BIBentrySTDinterwordspacing

\bibitem{Dehak:11}
N.~{Dehak}, P.~J. {Kenny}, R.~{Dehak}, P.~{Dumouchel}, and P.~{Ouellet},
  ``Front-end factor analysis for speaker verification,'' \emph{IEEE
  Transactions on Audio, Speech, and Language Processing}, vol.~19, no.~4, pp.
  788--798, 2011.

\bibitem{Shum:13}
S.~H. {Shum}, N.~{Dehak}, R.~{Dehak}, and J.~R. {Glass}, ``Unsupervised methods
  for speaker diarization: An integrated and iterative approach,'' \emph{IEEE
  Transactions on Audio, Speech, and Language Processing}, vol.~21, no.~10, pp.
  2015--2028, 2013.

\bibitem{Variani:14}
E.~{Variani}, X.~{Lei}, E.~{McDermott}, I.~L. {Moreno}, and
  J.~{Gonzalez-Dominguez}, ``Deep neural networks for small footprint
  text-dependent speaker verification,'' in \emph{2014 IEEE International
  Conference on Acoustics, Speech and Signal Processing (ICASSP)}, 2014, pp.
  4052--4056.

\bibitem{xvector}
\BIBentryALTinterwordspacing
D.~{Snyder}, D.~{Garcia-Romero}, G.~{Sell}, D.~{Povey}, and S.~{Khudanpur},
  ``X-vectors: Robust {DNN} embeddings for speaker recognition,'' in \emph{2018
  IEEE International Conference on Acoustics, Speech and Signal Processing
  (ICASSP)}, 2018, pp. 5329--5333. [Online]. Available:
  \url{https://ieeexplore.ieee.org/document/8461375}
\BIBentrySTDinterwordspacing

\bibitem{Wan:18}
L.~{Wan}, Q.~{Wang}, A.~{Papir}, and I.~L. {Moreno}, ``Generalized end-to-end
  loss for speaker verification,'' in \emph{2018 IEEE International Conference
  on Acoustics, Speech and Signal Processing (ICASSP)}, 2018, pp. 4879--4883.

\bibitem{Wang:18}
Q.~{Wang}, C.~{Downey}, L.~{Wan}, P.~A. {Mansfield}, and I.~L. {Moreno},
  ``Speaker diarization with {LSTM},'' in \emph{2018 IEEE International
  Conference on Acoustics, Speech and Signal Processing (ICASSP)}, 2018, pp.
  5239--5243.

\bibitem{Mansfield:18}
P.~A. Mansfield, Q.~Wang, C.~Downey, L.~Wan, and I.~L. Moreno, ``Links: A
  high-dimensional online clustering method,'' \emph{ArXiv preprint}, vol.
  1801.10123, 2018.

\bibitem{Singh2020}
\BIBentryALTinterwordspacing
P.~Singh and S.~Ganapathy, ``Deep self-supervised hierarchical clustering for
  speaker diarization,'' in \emph{Proceedings of Interspeech 2020}, 2020, pp.
  294--298. [Online]. Available:
  \url{http://dx.doi.org/10.21437/Interspeech.2020-2297}
\BIBentrySTDinterwordspacing

\bibitem{Zhang:19}
A.~{Zhang}, Q.~{Wang}, Z.~{Zhu}, J.~{Paisley}, and C.~{Wang}, ``Fully
  supervised speaker diarization,'' in \emph{2019 IEEE International Conference
  on Acoustics, Speech and Signal Processing (ICASSP)}, 2019, pp. 6301--6305.

\bibitem{scikit-learn}
F.~Pedregosa, G.~Varoquaux, A.~Gramfort, V.~Michel, B.~Thirion, O.~Grisel,
  M.~Blondel, P.~Prettenhofer, R.~Weiss, V.~Dubourg, J.~Vanderplas, A.~Passos,
  D.~Cournapeau, M.~Brucher, M.~Perrot, and E.~Duchesnay, ``Scikit-learn:
  Machine learning in {P}ython,'' \emph{Journal of Machine Learning Research},
  vol.~12, pp. 2825--2830, 2011.

\bibitem{Hinton:03}
\BIBentryALTinterwordspacing
G.~E. Hinton and S.~T. Roweis, ``Stochastic neighbor embedding,'' in
  \emph{Advances in Neural Information Processing Systems 15}, S.~Becker,
  S.~Thrun, and K.~Obermayer, Eds.\hskip 1em plus 0.5em minus 0.4em\relax MIT
  Press, 2003, pp. 857--864. [Online]. Available:
  \url{http://papers.nips.cc/paper/2276-stochastic-neighbor-embedding.pdf}
\BIBentrySTDinterwordspacing

\bibitem{Scudder:65}
H.~{Scudder}, ``Probability of error of some adaptive pattern-recognition
  machines,'' \emph{IEEE Transactions on Information Theory}, vol.~11, no.~3,
  pp. 363--371, 1965.

\bibitem{Yarowsky:95}
\BIBentryALTinterwordspacing
D.~Yarowsky, ``Unsupervised word sense disambiguation rivaling supervised
  methods,'' in \emph{33rd Annual Meeting of the Association for Computational
  Linguistics}.\hskip 1em plus 0.5em minus 0.4em\relax Cambridge,
  Massachusetts, USA: Association for Computational Linguistics, Jun. 1995, pp.
  189--196. [Online]. Available:
  \url{https://www.aclweb.org/anthology/P95-1026}
\BIBentrySTDinterwordspacing

\bibitem{vizier}
\BIBentryALTinterwordspacing
D.~Golovin, B.~Solnik, S.~Moitra, G.~Kochanski, J.~Karro, and D.~Sculley,
  ``Google vizier: A service for black-box optimization,'' in \emph{Proceedings
  of the 23rd ACM SIGKDD International Conference on Knowledge Discovery and
  Data Mining}, ser. KDD '17.\hskip 1em plus 0.5em minus 0.4em\relax New York,
  NY, USA: ACM, 2017, pp. 1487--1495. [Online]. Available:
  \url{http://doi.acm.org/10.1145/3097983.3098043}
\BIBentrySTDinterwordspacing

\bibitem{Sell:15}
G.~{Sell} and D.~{Garcia-Romero}, ``Diarization resegmentation in the factor
  analysis subspace,'' in \emph{2015 IEEE International Conference on
  Acoustics, Speech and Signal Processing (ICASSP)}, 2015, pp. 4794--4798.

\bibitem{Bredin2017}
\BIBentryALTinterwordspacing
H.~Bredin, ``pyannote.metrics: A toolkit for reproducible evaluation,
  diagnostic, and error analysis of speaker diarization systems,'' in
  \emph{Proceedings of Interspeech 2017}, 2017, pp. 3587--3591. [Online].
  Available: \url{http://dx.doi.org/10.21437/Interspeech.2017-411}
\BIBentrySTDinterwordspacing

\bibitem{Castaldo:08}
F.~{Castaldo}, D.~{Colibro}, E.~{Dalmasso}, P.~{Laface}, and C.~{Vair},
  ``Stream-based speaker segmentation using speaker factors and eigenvoices,''
  in \emph{2008 IEEE International Conference on Acoustics, Speech and Signal
  Processing (ICASSP)}, 2008, pp. 4133--4136.

\bibitem{Senoussaoui:14}
M.~{Senoussaoui}, P.~{Kenny}, T.~{Stafylakis}, and P.~{Dumouchel}, ``A study of
  the cosine distance-based mean shift for telephone speech diarization,''
  \emph{IEEE/ACM Transactions on Audio, Speech, and Language Processing},
  vol.~22, no.~1, pp. 217--227, 2014.

\bibitem{Garcia-Romero:17}
D.~{Garcia-Romero}, D.~{Snyder}, G.~{Sell}, D.~{Povey}, and A.~{McCree},
  ``Speaker diarization using deep neural network embeddings,'' in \emph{2017
  IEEE International Conference on Acoustics, Speech and Signal Processing
  (ICASSP)}, 2017, pp. 4930--4934.

\end{thebibliography}

\end{document}